# TO THE VACUUM CONDITIONS THAT PROVIDED THE PROCESS OF ELEKTROMAGNETIC WAVES GENERATED BY A RELATIVISTIC MAGNETRON

*A.B. Batrakov, S.I. Fedotov, O.M. Lebedenko, I.N. Onishchenko, O.L. Rak,*
*M.V. Volovenko, Yu.N. Volkov*
*National Science Center «Kharkiv Institute of Physics and Technology»,*
*Kharkiv, Ukraine*
*E-mail: a.batrakov67@gmail.com*

The powerful EHF radiation generated in a relativistic high-voltage pulsed magnetron of the 8 mm range is experimentally investigated. The factors that negatively affect the generation of EHF radiation are experimentally investigated and analyzed. It is determined that the processes that reduce the efficiency and duration of the generation pulse also include low pressure in the vacuum diode of the magnetron. It is known that the main residual gases in the vacuum diode of a magnetron are air components. During magnetron operation, there is a significant increase in the pressure of the residual atmosphere: hydrocarbons, water vapor, and hydrogen. A vacuum system was designed to evacuate these gases to ensure optimal operation of the magnetron. For the new vacuum system, a cryogenic condensation-adsorption pump was developed and applied, which allowed to increase the rate of pumping out the main residual gases. The peculiarity of the developed pump is that the pumping element with the adsorbent has a working surface that is twice as large as the one in the corrugated form of the sorption cartridge. Another feature of the pump is its efficiency, which is achieved through the use of nitrogen vapor for cooling the inter-wall gap. The use of cryogenic pumping means made it possible to obtain a pressure of $1·10^{-6}$ Torr, which led to a 25 percent increase in the EHF radiation of the relativistic magnetron.

## INTRODUCTION

The use of high-energy EHF radiation is necessary in a variety of applications, including plasma physics and many fields of applied research [1]. In the millimeter wavelength range, researchers of relativistic magnetrons face problems with the efficiency of EHF energy output, both in the radial direction of the cylindrical structure and in the axial direction [2]. Such undesirable effects as an increase in the duration of the voltage in the electrodynamic structure (ESD), and a pressure of $1·10^{-4}$ Torr led to a rapid deterioration of the ESD [3-6]. Therefore, the question arose of improving the method of high-frequency power dissipation from the magnetron's EMF [7-9]. One of the steps that contributes to this is to obtain a residual gas pressure of $1·10^{-6}$ Torr. It is known that in the vacuum diode of a magnetron, the main residual gases are air components. During magnetron operation, there is a significant increase in the residual atmosphere: hydrocarbons, water vapor, and hydrogen. Prior to the modernization, the vacuum pumping system consisted of a fore-vacuum pump for low vacuum and a magnet discharge pump for high vacuum of $1·10^{-4}$-$6·10^{-5}$ Torr. Such a configuration of vacuum pumping equipment was not good for pumping hydrocarbons, hydrogen, and water vapor. This is one of the reasons why the magnetron operated with low efficiency. At the same time, the working surfaces of the discharge pump magnet were out of order, which led to shutdowns. To generate EHF radiation in the magnetron, it is necessary to maintain a certain pressure in the vacuum system and it should be $1·10^{-6}$ Torr. This is the goal of the work. Achieving the deepest vacuum provides optimal conditions for the formation and stable maintenance of the electron flux interacting with the electromagnetic field. This paper presents the results of increasing EHF radiation by 25% by obtaining an operating vacuum of $1·10^{-6}$ Torr.

## VACUUM SYSTEM OF A MAGNETRON

The developed vacuum circuit provides optimal conditions for the formation and stable maintenance of the electron flux interacting with the electromagnetic field. The increase of EHF radiation is proposed by obtaining a pressure in the vacuum system of the magnetron at the level of $P=1·10^{-6}$ Torr. The studies were carried out on the installation, the scheme of which is shown in Fig. 1.

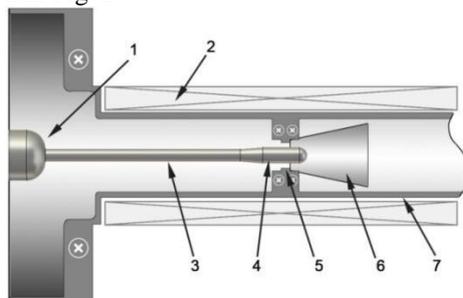

*Fig.1. Schema of a relativistic magnetron*
*1. Vacuum resonator. 2. Solenoid. 3. Cathode holder. 4. Cathode. 5. Anode. 6. The output horn.*
*7. Supersized waveguide.*

Experiments with the RM-48 magnetron revealed the EHF generation at a frequency of 36 to 41 GHz with a magnetic induction value of $B_0=0,35$ to $0,8$ T and an anode voltage of $U_0=190$ kV to $250$ kV. The diameters of the anode and cathode were 25 mm and 14 mm, respectively, and the gap between the cathode and anode was 3 mm to 5 mm. The observed oscillations can be defined as $\pi/2$; $\pi$ or $(2/3)\pi$ modes. Nevertheless, the final result of the experiments with axial EHF power dissipation with horn support still needs to be classified as unsatisfactory.

One of the reasons for this problem is the influence of the anode-cathode plasma on the discharge. The worse the vacuum conditions in the magnetron, the greater the influence of the anode-cathode plasma on the discharge. Several problems can occur if the vacuum level in the magnetron is not maintained properly:

1. Electron flux instability: Insufficient vacuum can lead to fluctuations and instability in the electron flux, which will negatively affect the generation of the EHF radiation.

2. Increased gas pressure: At elevated gas pressures in the vacuum chamber, unwanted electron collisions with gas molecules can occur, reducing the efficiency of the magnetron.

3. Overheating of elements: A low vacuum can cause the magnetron and other system components to overheat, which can cause damage or failure.

4. Reduced service life: Constant fluctuations in vacuum levels can shorten the life of the magnetron and other system components.

5. Breakdown between the cathode and anode: The literature shows that the pressure in the magnetron vacuum chamber deteriorates by an order of magnitude [10].

The studies conducted suggest that a relativistic magnetron can produce a prebreakdown in the anode-cathode gap. The current that appears in this case is 50-100 μA. In this case, we see significant destruction of the anode, which looks like potholes. The state of the anode and cathode surfaces is shown in Figure 2.

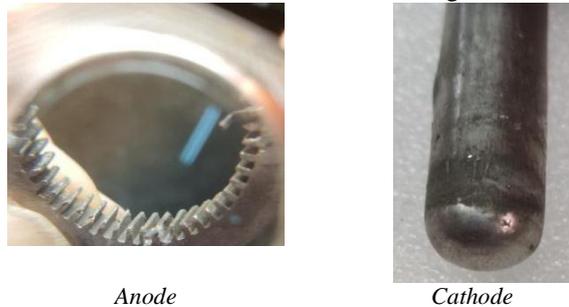

*Anode*      *Cathode*

*Fig. 2. Condition of the damaged surface of the anode and cathode of a relativistic magnetron*

The destruction at the anode indicates the electronic nature of the current. We also see the transfer of metal from the cathode to the anode. Metal is destroyed on the anode and soot appears. This is due to the inability to achieve the required vacuum level with the available pumping equipment. Therefore, the question arose of improving the vacuum system of the magnetron.

The new vacuum system was designed based on the requirements for optimal operation of the relativistic magnetron. The vacuum evacuation system includes: a fore-vacuum evacuation line, a high vacuum evacuation line. Fig. 3 shows a schematic of the vacuum evacuation system of the relativistic magnetron chamber.

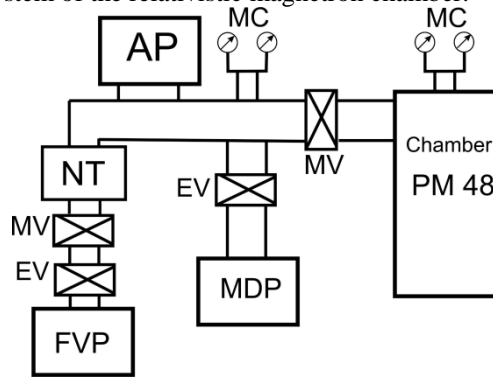

*Fig. 3. Diagram of the vacuum pumping system of the relativistic magnetron chamber.*
*FVP - fore-vacuum pump; AP - adsorption pump; MDP - magneto-dischargeble pump; MV - manually operated valves; EAV - electromechanically operated valves; PM48 Chamber - magnetron chamber; MC - pressure transducers; NT - nitrogen trap; EV - emergency vent valve.*

The vacuum system of the chamber, shown in Fig. 3, has a fore-vacuum evacuation line, which is designed to evacuate the Chamber PM48 from atmospheric pressure to $\sim 7 \cdot 10^{-2}$ Torr. The evacuation is performed by a fore-vacuum pump FVP through an open valve with an electromechanical drive EV and a nitrogen trap NT-LAF-32 filled with nitrogen. The pressure in the line and the working chamber is measured by the MC pressure transducers. The high-vacuum pumping line consists of an adsorption pump AP with a pressure transducer MC connected to the magnetron chamber Chamber PM48 through a manually operated valve MV, and a magnet discharge pump MDP connected to the relativistic magnetron chamber through an electromechanically operated valve EAV. The vacuum system also includes an emergency air inlet valve EV, which is activated in case of a power outage.

To obtain a working pressure in a relativistic magnetron at the level of $1·10^{-6}$ Torr, it is proposed to use cryogenic pumping means [11, 12]. For this purpose, a design of a high-vacuum cryogenic condensation-adsorption pump operating on a single refrigerant with efficient use of waste vapor was developed. The use of nitrogen vapor will increase the service life of the pump without increasing refrigerant consumption. The schematic of the developed cryogenic condensing-adsorption pump is shown in Fig. 4.

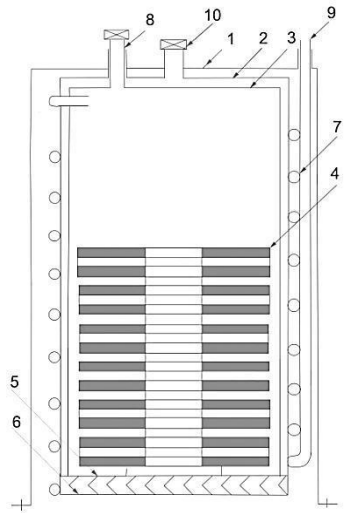

*Figure 4. Diagram of the condensing-adsorption pump*

The pump consists of an outer casing (1), which houses the pumping element in the form of an inner casing (3) with refrigerant and a screen (2). In the lower part of the inner casing (3), there are cassettes with adsorbent (4) (birch activated carbon). The bulk of the gas is pumped out by condensing on the condensing surface (6) of the inner casing (3). The non-condensed part of the gas is absorbed by the adsorbent after repeated contact with surfaces having the refrigerant temperature. The screen (2) with chevron (6) is designed to protect the inner casing (3) from external radiation and other thermal flows. The screen (2) is cooled by steam flowing through the coil (7). It also freezes the components condensing from the pumped volume. The surfaces of the inner casing (3) with the refrigerant and the elements of the screen (2) are wrapped in mylar to minimize the heat inflow to them. The screen (2) and the inner casing (3) with the refrigerant are suspended in the outer casing (2) of the pump on two nozzles (8) and (9) so that it is possible to fill the refrigerant through nozzle (8) and to remove refrigerant vapors through nozzle (9). To reduce the thermal load on the inner casing, the volume between the inner casing (3) and the screen (2) is pumped out through the nozzle (10).

The vacuum system operates as follows:

1. With the MV1, MV2, MV3, EAV valves open, we pump out the accelerator chamber with the EVP AVZ-20 fore-vacuum pump through the NT LAF-32 nitrogen trap filled with liquid nitrogen to a pressure of $7·10^{-2}$ Torr, which is the transitional mode between the viscous and molecular modes of gas flow. After that, we close the MV1 valve and turn off the fore-vacuum pump.

2. Close the MV3 valve and pour nitrogen into the adsorption pump AP, after starting the AP, open the MV3 valve and pump the system to $3·10^{-4}$ Torr.

3. Close the EAV valves and prepare the MDP magnetic discharge pump for operation. Having obtained a pressure of $2·10^{-4}$ Torr, we open the MDP pump to the magnetron chamber. This allows us to obtain a pressure of $1·10^{-6}$ Torr.

The vacuum system we have modernized allows the relativistic magnetron to operate in a cyclic mode without failure of the pumping elements.

## EXPERIMENTAL PART

After modernization of the vacuum system of the relativistic magnetron, experiments were carried out to obtain EHF radiation at different residual atmospheric pressures in the vacuum volume of the magnetron. The magnetic induction was $B_0=0.75$ T, the anode voltage $U_0=250$ kV. The output EHF radiation from the magnetron passed through a circular waveguide with a diameter of 80 mm and a length of 775 mm and was output to free space through an organic glass window. The output power of the emitted EHF was estimated by measuring the spatial distribution of EHF from the magnetron. The receiving antenna was a pyramidal horn with an open end ($4\times3.2$ cm$^2$), which was placed in the azimuthal direction at a distance of 1.35 m from the window. The envelope of the output EHF signal was obtained using a D404 diode and recorded by a digital oscilloscope with a bandwidth of 200 MHz. The first experiment was carried out at a pressure in the vacuum volume of the magnetron of $P=3\cdot10^{-4}$ Torr, the second at a pressure of $P=1\cdot10^{-6}$ Torr. The voltage, current, and EHF radiation oscillograms are shown in Fig. 5.

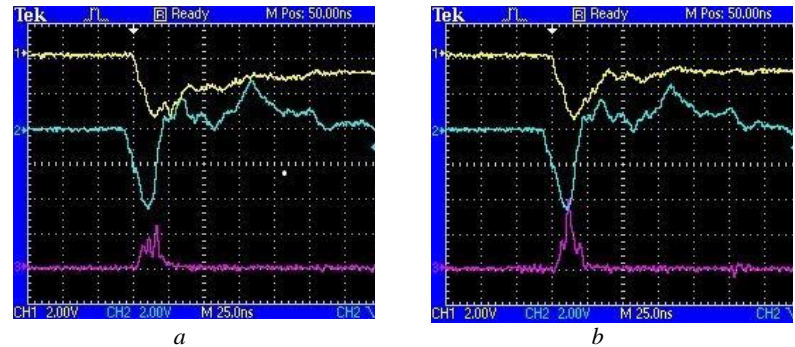

*a*           *b*

*Fig. 5. Voltage, current and EHF radiation waveforms*
*a- waveform at the pressure in the vacuum volume of the magnetron $P= 3\cdot10^{-4}$ Torr*
*b- waveform at the pressure in the vacuum volume of the magnetron $P= 1\cdot10^{-6}$ Torr*

In the waveforms, the first ray is the anode voltage, the second ray is the current, and the third ray is the EHF radiation.

The calculated EHF radiation power in the first case was 650 kW, and in the second case 875 kW. Obtaining the pressure of residual gases in the relativistic magnetron at the level of $P = 1\cdot10^{-6}$ Torr gives an increase in EHF radiation power by 25 percent.

## CONCLUSION

The vacuum system presented allows to obtain an oil-free vacuum. It provides optimal conditions for the formation and stable maintenance of the electron flux interacting with the electromagnetic field. The vacuum evacuation system allows you to quickly reach an operating pressure of $1\cdot10^{-6}$ Torr, maintain the installation in the evacuated state for a long time and reduce the time for preparing the vacuum chamber between pulses. Obtaining the residual gas pressure in the relativistic magnetron at the level of $P= 1\cdot10^{-6}$ Torr gives an increase in EHF radiation power by 25 percent.

ORCID IDs

A.B. Batrakov http://orcid.org/0000-0001-6158-2129
S.I. Fedotov http://orcid.org/0000-0002-7216-0615
M.V. Volovenko http://orcid.org/0000-0001-7216-2058
O.M. Lebedenko http://orcid.org/0009-0004-2243-8393
I.N. Onishchenko http://orcid.org/0000-0002-8025-5825
Yu.N. Volkov http://orcid.org/0009-0002-0557-8090
O.L. Rak http://orcid.org/0009-0000-6683-1235

## ДО УМОВ ВАКУУМУ, ЩО ЗАБЕЗПЕЧУЮТЬ ПРОЦЕС ГЕНЕРАЦІЇ ЕЛЕКТРОМАГНІТНИХ ХВИЛЬ РЕЛЯТИВІСТСЬКИМ МАГНЕТРОНОМ

*A.B. Batrakov, S.I. Fedotov, O.M. Lebedenko, I.N. Onishchenko, O.L. Rak, M.V. Volovenko, Yu.N. Volkov*

Експериментально досліджено потужне НВЧ-випромінювання, що генерується в релятивістському високовольтному імпульсному магнетроні діапазону 8 мм. Експериментально досліджено та проведено аналіз факторів, що негативно впливають на генерацію НВЧ випромінювання. Визначено, що до процесів, котрі зменшують ефективність та тривалість імпульсу генерації, відноситься також низький тиск у вакуумному діоді магнетрону. Відомо, що в вакуумному діоді магнетрону головними залишковими газами є компоненти повітря. При роботі магнетрону спостерігається значне збільшення тиску залишкової атмосфери: вуглеводнів, водяної пари і водню. Для їх відкачування була скомпонована вакуумна система, що забезпечує оптимальну роботу магнетрону. Для нової вакуумної системи було розроблено та застосовано кріогенний конденсаційно-адсорбційний насос, якій дозволив збільшити швидкість відкачування основних залишкових газів. Особливість розробленого насосу полягає в тому, що відкачувальний елемент з адсорбентом має збільшену в два рази робочу поверхню-за рахунок використання гофріванової форми сорбційного патрона. Іншою особливістю насосу є його економічність, вона досягнута за рахунок використання парів азоту для охолодження між стінного проміжку. За рахунок використання кріогенних засобів відкачування вдалося отримати тиск на рівні $1\cdot10^{-6}$ Тор, що призвело до збільшення випромінювання НВЧ релятивістського магнетрону на рівні 25 відсотків.